\pdfoutput=1
\documentclass[12pt,a4paper]{article}
\usepackage[utf8]{inputenc}
\usepackage{amsmath,amssymb,amsfonts,graphicx}
\usepackage[square,numbers,sort&compress]{natbib}
\usepackage[colorlinks,allcolors=blue]{hyperref}

\newcommand{\beq}{\begin{eqnarray}}
\newcommand{\eeq}{\end{eqnarray}}
\newcommand{\non}{\nonumber\\}
\DeclareMathOperator{\U}{U}
\DeclareMathOperator{\SU}{SU}

\newcommand{\p}{\partial}
\renewcommand{\i}{\mathrm{i}}
\renewcommand{\d}{\mathop{}\!\mathrm{d}}

\newcommand{\bbD}{\mathbb{D}}
\newcommand{\dA}{\delta{\mkern-2.5mu}A}

\DeclareMathOperator{\vol}{vol}

\DeclareMathOperator{\tr}{tr}
\DeclareMathOperator{\Tr}{Tr}
\DeclareMathOperator{\ind}{index}
\newcommand{\calA}{\mathcal{A}}
\newcommand{\calE}{\mathcal{E}}
\newcommand{\calF}{\mathcal{F}}

\newcommand{\Pm}{{\mathcal{P}_-}}
\newcommand{\Pp}{{\mathcal{P}_+}}

\makeatletter
\newcommand{\doublewidetilde}[1]{{%
  \mathpalette\double@widetilde{#1}%
}}
\newcommand{\double@widetilde}[2]{%
  \sbox\z@{$\m@th#1\widetilde{#2}$}%
  \ht\z@=.9\ht\z@
  \widetilde{\box\z@}%
}
\makeatother

\begin{document}
\pagenumbering{Roman}
\begin{titlepage}
  \begin{flushright}
    May 2023
  \end{flushright}
\vskip3cm
\begin{center}
  {\Large\bf Modes of the Sakai-Sugimoto soliton}\\[2cm]
  {\bf Markus A.~G.~Amano, Sven Bjarke
    Gudnason}\\[2cm]
{Institute of Contemporary Mathematics, School of Mathematics and Statistics, Henan University, Kaifeng, Henan 475004, P. R. China}
\end{center}
\vfill
\begin{abstract}
  The instanton in the Sakai-Sugimoto model corresponds to the
  Skyrmion on the holographic boundary -- which is asymptotically flat --
  and is fundamentally different from the flat Minkowski space Yang-Mills
  instanton.
  We use the Atiyah-Patodi-Singer index theorem and a series of
  transformations to show that there are $6k$ zeromodes -- or moduli
  -- in the limit of infinite 't Hooft coupling of the Sakai-Sugimoto
  model.
  The implications for the low-energy baryons -- the Skyrmions -- on
  the holographic boundary, is a scale separation between $2k$
  ``heavy'' massive modes and $6k-9$ ``light'' massive modes for $k>1$; the $9$
  global transformations that correspond to translations, rotations
  and isorotations remain as zeromodes. For $k=1$ there are 2
  ``heavy'' modes and 6 zeromodes due to degeneracy between rotations
  and isorotations.
\end{abstract}
\vfill
\rule{10cm}{0.5pt}\\
{\tt magarbiso@crimson.ua.edu}, {\tt gudnason@henu.edu.cn}
\end{titlepage}

\pagenumbering{arabic}
\section{Introduction}

The Witten-Sakai-Sugimoto (WSS) model \cite{Witten:1998zw,Sakai:2004cn} is
perhaps the most popular top-down holographic model of QCD to date,
see Ref.~\cite{Rebhan:2014rxa} for a review.
The model is a string-theory construction of QCD, which at low
energies reduces to a Yang-Mills action with a Chern-Simons term,
describing the quark flavor degrees of freedom with the gauge dynamics
encoded in the background metric of a cigar shaped AdS-like manifold.
The low-energy effective action of the latter yields the Skyrme model
coupled to an infinite tower of vector mesons.
The baryon in the bulk of the 5-dimensional AdS-like spacetime is an
instanton, known as the Sakai-Sugimoto (SS) soliton
\cite{Bolognesi:2013nja}, which corresponds to the Skyrmion coupled to
an infinite tower of vector mesons on the boundary -- the conformal
boundary being a 4-dimensional flat spacetime.

The topology of the instanton or baryon in the WSS model can be viewed
as a map from a large 3-sphere at ``infinity'' of a timeslice of the
5D bulk, to SU(2). Such maps are characterized by
$\pi_3(S^3)=\mathbb{Z}\ni k$, where $S^3$ is the manifold
corresponding to SU(2) and $k$ is the instanton or baryon number.
Holographically, the instanton number is also the Skyrmion or baryon
number on the boundary, where the topology instead is given by a map
of a timeslice of the entire conformal boundary with the point at
infinity, i.e.~$\mathbb{R}^3\cup\{\infty\}\simeq S^3$, which is also a
3-sphere, still mapping to SU(2).

When the 't Hooft coupling is very large, the Yang-Mills action
dominates over the Chern-Simons term, and it has been shown by
Bolognesi and Sutcliffe that the flat-space or self-dual instanton is
a good approximation to the SS soliton in that limit
\cite{Bolognesi:2013nja}.
For the 't Hooft coupling of order one, curvature effects of the
background AdS-like space come into play, making the approximation
unsuitable.

The moduli space of instantons on $\mathbb{R}^4$ has been studied in
quite some detail in the literature
\cite{Donaldson:1992,Dorey:2002ik}; they are concrete examples of
hyper-K\"ahler manifolds \cite{Hitchin:1986ea}.
Their dimension is determined by the Atiyah-Singer index theorem
\cite{Atiyah:1968mp}.
Nevertheless, it requires one to perform a conformal
transformation to go to the 4-sphere, so the index theorem for compact
manifolds can be applied.\footnote{Recent applications of the Atiyah-Patodi-Singer index theorem
\cite{Atiyah:1975jf} have been used to study the moduli of field
theories with boundaries \cite{Kobayashi:2021jbn,Matsuki:2021zct}.}
In particular, the (real) dimension of the instanton moduli space is $8k$, which
corresponds to the individual instanton having $4$ translational zeromodes, 
$3$ rotational zeromodes (inside SU(2)) and $1$ ``size'' zeromode.

The curved spacelike timeslice of the AdS-like spacetime
lifts two of these zeromodes per individual instanton. In
particular, the position in the holographic direction and the size of the
(individual) instantons are no longer moduli \cite{Hata:2007mb}. That is, the
SS solitons prefer to sit at the IR tip of the cigar with a size stabilized by
the competition of the electric flux coming from the Chern-Simons term and
gravity.

We can illustrate the loss of these two moduli in the WSS model, in the limit
of infinite 't Hooft coupling ($\lambda$), by choosing the flat space
instanton Ansatz 
\beq
A_I = -\sigma_{IJ}x_J b(\rho),
\eeq
with $I,J=1,2,3,4$, $\sigma_{IJ}$ is the antisymmetric 't Hooft
tensor with $\sigma_{ij}=\epsilon_{ijk}\sigma^k$,
$\sigma_{4i}=\sigma^i$, and $\rho$ is the radial coordinate
$\rho^2=x_I^2$ and finally, $b(\rho)$ is a profile function.
Although the Yang-Mills action is somewhat complicated with this
Ansatz, it simplifies drastically by plugging in the flat space
solution $b(\rho)=1/(\rho^2+\beta^2)$:
\beq
S_{\rm YM} = -24\kappa\int\d^5x\;\frac{\beta^4}{(\rho^2+\beta^2)^4}\big(k(x^4) + h(x^4)\big),
\eeq
with $k(z) = 1 + z^2$, and $h(z) = k^{-1/3}(z)$ are metric functions,
coming from the metric
\beq
g_5 = H(x^4)\d x_\mu\d x^\mu + \frac{1}{H(x^4)}(\d x^4)^2,\qquad
H(z) = (1+z^2)^{2/3},
\label{eq:g5_H}
\eeq
with $\mu=0,1,2,3$ and $\mu$ being raised by the flat Minkowski
metric with mostly positive signature and $\kappa$ is a
constant proportional to the 't Hooft coupling $\lambda$.
If the metric factors were constants, the ``size'' modulus $\beta$ and
the 4 translations are manifestly leaving the 4-dimensional integral
invariant. 
But due to the curvature of the AdS-like background, the dependence on
$\beta$ is induced, which can be seen by first integrating over $x^i$,
$i=1,2,3$:
\beq
S_{\rm YM} = -3\pi^2\kappa\int\d x^0\d x^4\;
\frac{\beta^4}{\big((x^4)^2+\beta^2\big)^{5/2}}
\big(k(x^4) + h(x^4)\big).
\eeq
Integrating now over $x^4$, we have
\beq
S_{\rm YM} = -8\pi^2\kappa\int\d x^0\;\left[1 + \frac{\beta^2}{6} + \mathcal{O}(\beta^4)\right].
\eeq
This illustrates that the instanton in this AdS-like background wants
to shrink to zero size since ``pointlike'' instantons configurations
are energetically preferred. Of course, this is a consequence of taking the limit
of vanishing Chern-Simons term, which otherwise would stabilize the instanton
size by inducing an electric flux.
It is also clear that a shift of the instanton in
$x^i\in\mathbb{R}^3$, $i=1,2,3$ leaves the integral unchanged, whereas
a shift in $x^4\to x^4+\ell$ of the instanton -- i.e.~in the field
Ansatz, but not in the metric -- changes the action integral as
\begin{align}
S_{\rm YM} &= -3\pi^2\kappa\int\d x^0\d x^4\;
\frac{\beta^4}{\big((x^4+\ell)^2+\beta^2\big)^{5/2}}
\big(k(x^4) + h(x^4)\big)\non
&= -8\pi^2\int\d x^0\;\left[1 + \frac{\beta^2}{6} + \frac{2\ell^2}{3}(1+\beta^2) + \mathcal{O}(\beta^4,\ell^4)\right],
\end{align}
which shows that for any size, the instanton prefers to sit at the tip
of the cigar, i.e.~at $x^4=0$ ($\ell=0$).

The number of moduli of the single instanton in the AdS-like
spacetime, is thus given by 3 spatial translations (corresponding to
$x^i\in\mathbb{R}^3$) as well as 3 rotations of the solution inside
SU(2), which correspond to $A_I\to UA_IU^{-1}$, with $U\in\SU(2)$.
The total number of moduli is thus 6.

\begin{figure}[!tp]
  \centering
  \includegraphics{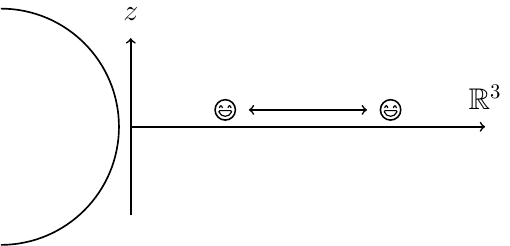}
\caption{\it Well separated point-like instantons sitting at the ``tip''
  of the cigar in the AdS-like spatial geometry of the WSS model. }
\label{fig:well_separated_instantons}
\end{figure}

The argument above is made for illustrative purposes and for a single
instanton.
Extending the argument to $k$ well separated instantons, is trivially
done by placing them at e.g.~$x^1=nL$, $L\gg\beta$, $L\gg1$,
$n=1,\cdots,k$, see Fig.~\ref{fig:well_separated_instantons}.
Of course, the calculation utilizes the flat space instanton solution
$b(\rho)=(\rho^2+\beta^2)^{-1}$, which is only an approximation for the
AdS-like spacetime of the WSS model.
The total number of moduli, which is also known as the dimension of
the moduli space, is thus
\beq
\dim_{\mathbb{R}}\mathcal{M}_k = 6k.
\eeq
In this paper, we will argue that the result holds in the WSS model,
not assuming any Ansatz or any large separation between the
instantons, but of course only in the $\lambda\to\infty$
limit, i.e.~at infinite 't Hooft coupling.

As for applications of this result, we will argue in the discussion of
the paper, that the $2k$ lifted moduli correspond to massive (or
vibrational) modes of Skyrmions and are parametrically set by the mass
scale $\lambda M_{\rm KK}\sim\lambda/R$, i.e.~the 't Hooft coupling times
the curvature radius of the AdS-like spacetime\footnote{The scalar
curvature, more precisely, is given by
$-\frac{16(3+4z^2)}{9(1+z^2)^{4/3}}\in\big[-\frac{16}{3},0\big]$,
which is (minus) an order one
number near $z\approx0$ in dimensionless variables. The curvature radius is thus an order one
number times $1/M_{\rm KK}$ with units restored.}.
On the other hand, of the $6k$ remaining moduli, $6k-9$ (with $k>1$) are lifted due
to the non-BPS-ness of the instanton in the WSS model with a finite
't Hooft coupling (or nonvanishing Chern-Simons level). These massive
or vibrational modes of the Skyrmion are parametrically sitting at the
mass scale $M_{\rm KK}\sim1/R$.

In the next section, we will study a version of the
Atiyah-Patodi-Singer index theorem for the instanton at hand.

\section{The index theorem}

The Witten-Sakai-Sugimoto (WSS) model is given by the Yang-Mills and
Chern-Simons actions
\begin{align}
  S &= \kappa\tr\int_{M_5}\calF\wedge \star\calF
  + \frac{9\kappa}{\lambda}\tr\int_{M_5}\omega_5,\\
  -\i\,\omega_5 &= \calA\wedge\calF^2
    -\frac12\calA^3\wedge\calF
    -\frac{1}{10}\calA^5,
\end{align}
where $\kappa=\frac{N_c\lambda}{216\pi^2}$ is an overall factor, $\lambda$ is
the 't Hooft coupling,
\beq
\calF
=\frac12(\p_\Gamma\calA_\Delta-\p_\Delta\calA_\Gamma
+[\calA_\Gamma,\calA_\Delta])\d x^\Gamma\wedge\d x^\Delta,
\eeq
is the field strength in 5 dimensions
($\Gamma,\Delta=0,1,2,3,4$),
$\calA_\Gamma=\calA_\Gamma^i\frac{\sigma^i}{2i}+\calA_\Gamma^0\frac{\mathbf{1}_2}{2i}$ is the U(2) gauge field, 
$\omega_5$ is the 5-dimensional Chern-Simons form,
and the powers of forms are understood by the wedge product.
The Hodge star $\star$ requires the manifold $M_5$ be equipped with a metric,
which is given in Eq.~\eqref{eq:g5_H}.\footnote{The following arguments can also be applied
for $H(x^4)=(1+(x^4)^2)^p$ where $p>1/2$. This is because the upcoming
$\xi$ coordinate is in a finite interval for $p>1/2$ (the interval
size depends on $p$ though). For the manifold to be AdS-like
  (possessing conformal boundary and the negative and finite curvature) $p\in
(1/2, 1]$ must be satisfied \cite{Bolognesi:2013nja}. }
The volume form on $M_5$ is
\beq
\vol_5 = \sqrt{-\det g_5}\d x^{01234} = \big(H(x^4)\big)^{3/2}\d x^{01234},
\eeq
with
$\d x^{01234}=\d x^0\wedge\d x^1\wedge\d x^2\wedge\d x^3\wedge\d x^4$
and $H$ is defined in Eq.~\eqref{eq:g5_H}.

First we take the large 't Hooft coupling limit ($\lambda\to\infty$),
so we are left with only the Yang-Mills action in 5 dimensions.
Without the Chern-Simons term in the action, the $\U(1)$ part of the
Yang-Mills action decouples, and we will concentrate on only the
$\SU(2)$ part in the remainder of this section.
Since we are only interested in the time-independent properties of the
instanton, we pick out a time slice of $M_5$ in order to work
with a 4-manifold. We do this by considering the embedding map from
$M_4$ to $M_5$:
\beq
\gamma : (x^1,x^2,x^3,x^4)\mapsto(0,x^2,x^3,x^4,x^1),
\eeq
which induces the metric on the 4-manifold $M_4$ via the pullback of
the metric $g_5$ (Eq.~\eqref{eq:g5_H}) by the map $\gamma$:
i.e.~$g_4=\gamma^*g_5$ yielding:
\beq
g_4 = \frac{1}{H(x^1)}(\d x^1)^2 + H(x^1)\d x^i\d x^i, \qquad i=2,3,4,
\eeq
where, for convenience, we have flipped the order of the ``holographic
direction'' from $x^4$ to $x^1$ -- this is just a relabeling done by
the map $\gamma$.
Although $g_5$ is a pseudo-Riemannian (Lorentzian) metric, the induced
metric $g_4$ is Riemannian. 
The time-independent action on $M_4$ is thus
simply
\beq
E = -\tr\int_{M_4} F\wedge *F,
\eeq
with $A=\gamma^*\calA$ and correspondingly $F=\gamma^*\calF$,
i.e.~they are the pullbacks of the gauge field and field strength,
respectively, and the Hodge star * is defined on $M_4$ (as opposed to
$\star$ that is defined on $M_5$). 
By the standard Bogomol'nyi completion, we can minimize the action $E$
on $M_4$ by
\beq
E = -\tr\int_{M_4}\frac12(F \mp *F)\wedge *(F \mp *F) \pm F\wedge F,
\eeq
where we have used that $**=1$ for 2-forms on a 4-dimensional
Riemannian manifold ($M_4$).
Since the first term is quadratic and the last term is the second
Chern class, i.e.~it is topological, the action $E$ is extremized
(minimized) by $F$ satisfying the (anti-)self-dual equation
\beq
F = \pm *F,
\eeq
where the upper sign is for self-dual fields and corresponds in
physics to instantons, whereas the lower sign is for anti-self-dual
fields and they correspond to anti-instantons.
From now on, let us fix the sign to the upper sign and work with
the self-dual equation:
\beq
\Pm F = \frac12(1-*)F = 0.
\label{eq:sd}
\eeq
$\Pm$ and $\Pp$ are the projection operators of 2-forms onto the self-dual and
anti-self-dual 2-forms, respectively.
The second Chern class of $A$ is given by
\begin{align}
  c_2 = -\frac{1}{8\pi^2}\tr\int_{M_4}F\wedge F = k,
  \label{eq:c2}
\end{align}
where $k$ is the number of instantons.
Hence, the action $E$ is bounded from below by the BPS bound
\beq
E \geq 8\pi^2k,
\eeq
which is saturated by (anti-)self-dual field configurations.

Considering now linear perturbations of the self-dual equation
\eqref{eq:sd}, we obtain
\beq
\d_A\dA - *\d_A\dA = 0,
\label{eq:fluc}
\eeq
where $\d_A=\d+[A,\circ]$ is the gauge covariant derivative with the
background gauge field $A$.
It is necessary to fix the gauge of the fluctuation fields such that
the fluctuation fields are themselves not gauge transformations
\cite{Atiyah:1978wi}, which we can do by the so-called Lorentz gauge
condition
\beq
\d_A^\dag\dA = -*\d_A*\dA=0,
\label{eq:Lorentzgauge}
\eeq
where $\d_A^\dag$ is the gauge covariant coderivative.
At this point, we could in principle combine the three independent
first order equations of Eq.~\eqref{eq:fluc} and that of
Eq.~\eqref{eq:Lorentzgauge}, but due to the complicated metric it will
prove to be a complicated computation. The form of the resulting
operator is relatively similar, even in this coordinate
basis/conformal class representation.

Therefore, we will first make a change of the holographic coordinate
\beq
\frac{\d z^2}{H(z)} = H(z)\d\xi^2.
\eeq
Choosing the positive root and changing variables, we arrive at the
conformally flat metric
\beq
\tilde{g} = H(\xi)\left(\d\xi^2 + \d x^i\d x^i\right),\qquad i=2,3,4,
\eeq
where $H(\xi)$ is given in terms of the standard
hypergeometric function
\beq
\xi = {}_2F_1\left(\tfrac12,\tfrac23;\tfrac32;-z^2\right)z.
\label{eq:xi_coords}
\eeq
\begin{figure}[!tp]
  \centering
  \includegraphics{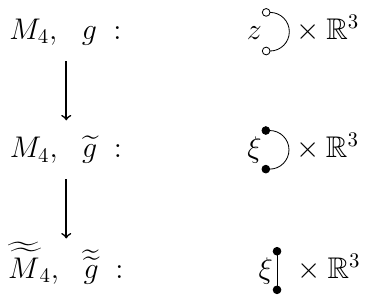}
\caption{\it The coordinate transformation from $z$ to $\xi$, followed by
  the conformal transformation to $\doublewidetilde{M}_4$, as
  described in the text. The open endpoints denote that the coordinate
  runs over an infinite range, whereas the closed (filled) endpoints
  denote a finite range. }
\label{fig:transformations}
\end{figure}
Now we will perform an active conformal transformation, since the
induced theory on $M_4$ is conformally invariant; this is possible
since we are only interested in the zeromodes of the theory.
It will however lead to nonstandard fall offs of the fields at the
would-be conformal boundary, which corresponds to $\xi\to\pm\pi$
($z\to\pm\infty$).
More precisely, on $M_4$ with coordinates corresponding to the metric
$\tilde{g}$, the holographic principle requires the fields at the
conformal boundary to fall off like
\beq
\lim_{|\xi|\to\pi} k(z(\xi))\p_iA_1 = {\rm const}.
\eeq
After the active conformal transformation, $\p_iA_1$ will tend to a
constant, instead of a polynomial fall off (polynomial in
$z\sim\tan\tfrac{\xi}{2}$, or more precisely the inverse function of
Eq.~\eqref{eq:xi_coords}).
Taking into account such boundary effects has been studied by
Atiyah-Patodi-Singer \cite{Atiyah:1975jf}, see also
Ref.~\cite{Jackiw:1977pu}.
We will denote the manifold $\doublewidetilde{M}_4$ after
performing this conformal transformation, see
Fig.~\ref{fig:transformations}.
This will amount to the index of the operator
\beq
\bbD\dA
:= \left(-\d x^1 \wedge \d^\dag_A + 2\iota_{\p_1} (\Pm \d_A) \right)\dA 
\equiv  \eta^I D_I\dA.
\eeq
The operator $\bbD$ is a first-order differential operator.
Naturally, the operator implied by Eqs.~\eqref{eq:Lorentzgauge} and
\eqref{eq:fluc} would lead to an operator
$\widehat\bbD: \Omega_1\rightarrow \Omega_2 \oplus \Omega_-$.
$\Omega_n$ is the space of gauge valued $n$-forms, and $\Omega_-$ is
the space of anti-self-dual forms.
Using a ``contraction'' operator
$C:\Omega_2 \oplus\Omega_- \rightarrow \Omega_1$ in composition with $\widehat\bbD$ the
operator $C\circ\widehat\bbD: \Omega_1 \rightarrow \Omega_1$ is
constructed.
The kernel of $C$ is empty since $\p_1:=\frac{\p}{\p x^1}$ is a nonvanishing vector
field; therefore, the kernel of $\bbD$ is the space of solutions that
obey Eqs.~\eqref{eq:Lorentzgauge} and \eqref{eq:fluc}.
The operator,
$\bbD$, can be shown to be a Fredholm operator 
which implies that its analytical index is written as
\beq
\ind(\bbD) = \dim\ker(\bbD) - \dim\ker(\bbD^\dag).
\eeq
According to Theorem (3.10) of Ref.~\cite{Atiyah:1975jf}
the index is also equal to the following topological index:
\begin{align}
  \ind(\bbD) &= -\frac{1}{\pi^2}\tr\int_{M_4} F\wedge F - \frac{h+\eta(0)}{2}\non
  &= 8k - \frac{h}{2},
  \label{eq:index_theorem310}
\end{align}
where we have used the second Chern class \eqref{eq:c2} and that
$\eta(0)$ vanishes for operators that possess a $\mathbb{Z}_2$
symmetry \cite{Atiyah:1975jf}, which $\bbD$ does -- chiral symmetry
(left and right spinors are exchangeable if the operator is viewed as
a Dirac operator).
$h$ on the other hand corresponds to the dimension of the kernel of
the boundary operator; writing the full operator $\bbD$ as
\begin{align}
  \bbD &= D_1 + \eta^i D_i\non
  &= \p_1 + \eta^i D_i, \qquad i=2,3,4,
\end{align}
where on the second line, we have chosen the gauge $A_1=0$ for the
background gauge fields\footnote{Note that the nonvanishing asymptotic
behavior of $F_{i1}\sim\p_iA_1$ at $\xi\to\pm\pi$ is not lost by this
gauge transformation, but is merely transformed to the other fields.}
and
\beq
\eta^1 = \mathbf{1}_4,\quad
\eta^2 =
\begin{pmatrix}
  \i\sigma^2 & 0\\
  0 & \i\sigma^2
\end{pmatrix},\quad
\eta^3 =
\begin{pmatrix}
  0 & \sigma^3\\
  -\sigma^3 & 0
\end{pmatrix},\quad
\eta^4 =
\begin{pmatrix}
  0 & \sigma^1\\
  -\sigma^1 & 0
\end{pmatrix}.
\eeq

$h$ is now the index of the operator $\eta^i D_i$ on the boundary $Y$,
where $X=Y\times I$, with $I$ being the interval: $\xi\in I$.
Since the boundary $Y$ are simply two copies of
$\mathbb{R}^3\cup\{\infty\}\sim S^3$, 
the index of $\eta^i D_i$ is counting the topological degree
$B\in\pi_3(S^3)$, being the number of would-be ``Skyrmions''.
Clearly $B=k$.
Now due to the presence of two conformal boundaries (the interval $I$
has two endpoints), the final result of the index reads
\begin{align}
  \ind(\bbD) &= 8k - \frac{2\times 2k}{2} = 6k,
\end{align}
where $2k=8k/4$, which can be found intuitively by noting that the
conformal boundary is directed in $1/4$ of the four dimensions, and
the factor of two is due to the conformal boundary appearing at both
$z\to\pm\infty$.

More rigorously, let's compute the index of the bulk operator $\bbD$,
which is technically performed on $\doublewidetilde{M}_4$ without a
boundary or with the double of it, so the boundary is absorbed into
the bulk of the manifold; the result is the second Chern class, which
is the first term of Eq.~\eqref{eq:index_theorem310}.
For this computation, we need the adjoint operator:
\beq
\bbD^\dag = \bar\eta^\mu D_\mu, \qquad
\bar\eta^1 = -\eta^1,\qquad
\bar\eta^i = \eta^i,\qquad i=2,3,4
\eeq
which is defined using the inner product
\beq\label{eq:innerprod}
(X,Y) := \int_{\doublewidetilde{M}_4} \tr(X^\dag Y)\vol,
\eeq
where $\tr$ is the matrix trace and $\vol$ is the volume form on
$\doublewidetilde{M}_4$, on which the metric is
$\tilde{\tilde{g}}=\delta$.
The inner product of Eq.~\eqref{eq:innerprod} is equivalent to the
Hodge inner product.

Now let's evaluate the bulk index
\begin{align}
  \ind(\bbD) &=
  \lim_{\varepsilon\to0}\Tr \left(e^{-\varepsilon^2\bbD^\dag\bbD} - e^{-\varepsilon^2\bbD\bbD^\dag}\right)\non
  &=\lim_{\varepsilon\to0}\frac{\varepsilon^2}{2}\Tr\left((\bbD^\dag\bbD)^2 - (\bbD\bbD^\dag)^2\right)+\cdots\non
  &=\frac{1}{8\pi^2}\int_{\doublewidetilde{M}_4}\tr\left[
    \left(-D_\mu D^\mu - {\epsilon_{1\rho}}^{\mu\nu}\eta^\rho F_{\mu\nu}\right)^2
    -\left(-D_\mu D^\mu\right)^2
    \right]\vol\non
  &=\frac{1}{8\pi^2}\int_{\doublewidetilde{M}_4}\tr\left[
    (\delta_1^{\mu}\bar\eta^\nu - \delta_1^{\nu}\bar\eta^\mu)F_{\mu\nu}{\epsilon_{1\rho}}^{\sigma\omega}\eta^\rho F_{\sigma\omega}
    \right]\vol\non
  &=-\frac{1}{4\pi^2}\int_{\doublewidetilde{M}_4}\tr\left[
    \epsilon^{\mu\nu\rho\sigma}F_{\mu\nu}F_{\rho\sigma}
    \right]\vol\non
  &=\frac{1}{8\pi^2}\int_{\doublewidetilde{M}_4}
    \epsilon^{\mu\nu\rho\sigma}F_{\mu\nu}^aF_{\rho\sigma}^a
    \vol\non
  &= 8k,
\end{align}
where in the second line, we have used that the gauge trace of the
linear order of the expanded exponentials vanishes
(i.e.~$\tr(\bbD^\dag\bbD)=\tr(\bbD\bbD^\dag)=0$) and that higher
than second orders vanish in the limit of $\varepsilon\to0$.
In the third line, we have used
\begin{align}
  \bbD^\dag\bbD &=
  (-\delta^{\mu\nu}
  -\delta_1^{\mu}\bar\eta^\nu
  +\delta_1^{\nu}\bar\eta^\mu
  -{\epsilon_{1\rho}}^{\mu\nu}\eta^\rho)D_\mu D_\nu\non
  &=-D_\mu D_\mu + \frac12(
  -\delta_1^{\mu}\bar\eta^\nu
  +\delta_1^{\nu}\bar\eta^\mu
  -{\epsilon_{1\rho}}^{\mu\nu}\eta^\rho)F_{\mu\nu}\non
  &=-D_\mu D_\mu - \eta^\rho\Pp_{\rho}^{\,~\mu\nu} F_{\mu\nu}\non
  &=-D_\mu D_\mu
  -{\epsilon_{1\rho}}^{\mu\nu}\eta^\rho F_{\mu\nu},\\
  \bbD\bbD^\dag &=
  (-\delta^{\mu\nu}
  +\delta_1^{\mu}\bar\eta^\nu
  -\delta_1^{\nu}\bar\eta^\mu
  -{\epsilon_{1\rho}}^{\mu\nu}\eta^\rho)D_\mu D_\nu\non
  &=-D_\mu D_\mu + \frac12(
  \delta_1^{\mu}\bar\eta^\nu
  -\delta_1^{\nu}\bar\eta^\mu
  -{\epsilon_{1\rho}}^{\mu\nu}\eta^\rho)F_{\mu\nu}\non
  &=-D_\mu D_\mu + \eta^\rho\Pm_{\rho}^{\,~\mu\nu} F_{\mu\nu}\non
  &=-D_\mu D_\mu,\label{eq:bbDbbDdag}
\end{align}
and inserted the normalizing prefactor that is composed by the
dimension of the $\eta$ matrices (i.e.~4), the group
Casimir, $C$, and finally the volume of the 3-sphere:
\beq
\frac{1}{4C\vol_{S^3}} = \frac{1}{4\frac122\pi^2} = \frac{1}{4\pi^2}.
\eeq
In the fourth line, we have used that the covariant Laplacians squared
cancel and that the cross product of the covariant Laplacian with the
field strength vanishes by the trace over $\eta^\rho$ due to
anti-symmetry of the Levi-Civita symbol.
Finally, we have used the property of the projection operator
\beq
\eta^\rho{\Pp_{\rho}}^{\mu\nu}F_{\mu\nu}
=\eta^\rho{\epsilon_{1\rho}}^{\mu\nu}F_{\mu\nu},\qquad
\eta^\rho{\Pm_{\rho}}^{\mu\nu}F_{\mu\nu} = 0,
\eeq
for self-dual background gauge fields.
In the fifth line, we have summed over $\mu$ and divided by four. The sixth
line is simply eight times the second Chern class in component form. The index
can be equated to the dimension of the kernel since the kernel of $\bbD^\dag$ is
empty. This is because $M_4$ is conformally equivalent to the non-negatively
curved manifold, $\doublewidetilde{M}_4$. This can be proven using a vanishing
theorem as was shown in Ref.~\cite{Atiyah:1978wi}.

Recalling that we work in the gauge $A_1:=0$ and hence the boundary
operator is $\p\bbD=\eta^iD_i$, $i=2,3,4$, $h$ is the dimension of the
kernel of
\beq
-D_iD_i - \frac12{\epsilon_{1k}}^{ij}\eta^kF_{ij},
\eeq
where we notice the factor of $\tfrac12$ is due to the ``electric''
part of the field strength not contributing to the boundary operator.
Since $\p\bbD$ is self-adjoint, we cannot compute the dimension of the
kernel via the same type of index theorem calculation.
Nevertheless, the dimension of the kernel of the operator can easily
be seen to be
\begin{align}
  \dim\ker(\p\bbD)
  &=\frac{1}{8\pi^2}\int_{\doublewidetilde{M}_4}\tr\left[
    \left(-\frac12{\epsilon_{1\rho}}^{\mu\nu}\eta^\rho F_{\mu\nu}\right)^2
    \right]\vol\non
  &= \frac{1}{32\pi^2}\int_{\doublewidetilde{M}_4}\tr\left[
    (\delta_1^{\mu}\bar\eta^\nu - \delta_1^{\nu}\bar\eta^\mu)F_{\mu\nu}{\epsilon_{1\rho}}^{\sigma\omega}\eta^\rho F_{\sigma\omega}
    \right]\vol\non
  &=-\frac{1}{16\pi^2}\int_{\doublewidetilde{M}_4}\tr\left[
    \epsilon^{\mu\nu\rho\sigma}F_{\mu\nu}F_{\rho\sigma}
    \right]\vol\non
  &=\frac{1}{32\pi^2}\int_{\doublewidetilde{M}_4}
    \epsilon^{\mu\nu\rho\sigma}F_{\mu\nu}^aF_{\rho\sigma}^a
    \vol\non
  &= 2k.
\end{align}
Using theorem \eqref{eq:index_theorem310}, we obtain the final result
for the number of zeromodes on $\doublewidetilde{M}_4$ with the
boundary effects:
\beq
\ind(\bbD) = 8k - 2k = 6k.
\eeq

\section{Discussion}

We have now obtained the number of zeromodes of the Sakai-Sugimoto
soliton or the instanton in a spatial slice of AdS$_5$ in the limit of
infinite 't Hooft coupling.
This limit is clearly not suitable for phenomenological studies of
baryons using holographic QCD, so what can we draw from this
computation?
The application of this result is a parametric dependence on the
scales in the system.
Let us write the (free) energy of the SS soliton as
\begin{align}
E &= \lambda\left(M_{\rm KK}\int\d^3x\d z\;s_{\rm YM} + \frac{M_{\rm KK}}{\lambda}\int\d^3x\d z\;s_{\rm CS}\right)\non
&= \lambda \calE(M_{\rm KK},\lambda),\label{eq:Efree}
\end{align}
with $s_{\rm YM}$ and $s_{\rm CS}$ being the dimensionless action densities of
the Yang-Mills and Chern-Simons parts of the theory, respectively.
Since $\calE(M_{\rm KK},\infty)$ does not depend on the 't Hooft coupling
$\lambda$, the $2k$ out of $8k$ modes that become massive have to be
at the mass scale $M_{\rm KK}$.
\begin{figure}[!tp]
  \centering
  \includegraphics{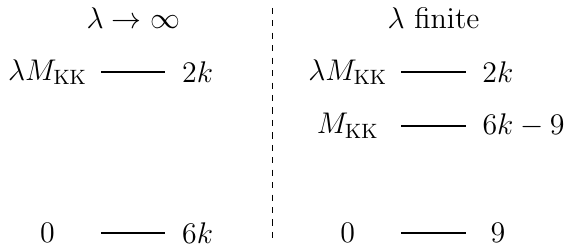}
\caption{\it The scales and number of modes at $\lambda\to\infty$
  ($\lambda$ finite) to the left (right) for $k>1$.}
\label{fig:scales}
\end{figure}
Now if we lower the 't Hooft coupling to a finite value, it is well
known that there are only 9 global zeromodes (translations,
rotations and isorotations) of the soliton with $k>1$
\cite{Braaten:1985np,Zahed:1986qz,Houghton:1997kg,Bolognesi:2013nja}; this
means that $6k-9$ massive modes are sitting at a
parametrically different scale, namely $M_{\rm KK}/\lambda$
(without the prefactor of $\lambda$ in Eq.~\eqref{eq:Efree}).
For $k=1$ on the other hand, there is a degeneracy between
rotations and isorotations due to spherical symmetry of the
solution, which means that there are 2 ``heavy'' modes and 6 zeromodes.
We summarize this conjecture in Fig.~\ref{fig:scales}.
The reason why this is a conjecture, is that we have not
computed the coefficients for each of the modes, which we assume to
be of order one.
Mathematically, the conjecture could only turn out to be wrong if
some of the coefficients of the modes were to vanish, which we know
from physics is not the case \cite{Bolognesi:2013nja}.

We note that once the 't Hooft coupling is lowered to a finite value,
the size of the SS soliton becomes finite \cite{Bolognesi:2013nja}.
Once this happens, the size and the curvature mixes and these massive
modes can be studied more in detail using spectral theory.
Rigorous bounds on the frequencies/masses of the modes would be an
interesting future direction.

Sutcliffe has proposed a model, which can be viewed as a flat
spacetime holographic model, where the scale of the theory is not
given by the curvature of the AdS-like background geometry, but
instead by the truncation of the number of vector bosons from an
infinite number to any finite number
\cite{Sutcliffe:2010et,Sutcliffe:2011ig}.
In that model, there is no Chern-Simons term and hence the 't Hooft
coupling cannot play a role as in the WSS model, since it is simply an
overall factor.
It would be interesting to consider the addition of the Chern-Simons
term in the Sutcliffe model, and see if it induces a scale separation
in that case too.

We note that the mechanism of breaking part of the zeromodes by having
a conformal boundary at $z\to\pm\infty$ is not present in the
Sutcliffe model, since it is defined on flat space: $\mathbb{R}^4$.
It could be interesting to study a crossover model, where one could
interpolate between having a conformal boundary and a truncation of
the number of vector bosons, to see how the zeromodes would be lifted
in such a hybrid model.

In the Skyrme model, which can be viewed as the low-energy effective
theory of the WSS model on the holographic boundary, in the limit
where all the infinite tower of vector bosons is decoupled, the
vibrational or massive modes of Skyrmions have been studied for baryon
numbers $B=1$ through $B=8$ \cite{BjarkeGudnason:2018bju}.
In Ref.~\cite{BjarkeGudnason:2018bju}, it was conjectured that the
Skyrmion possesses $7k$ ``light'' massive or vibrational
modes. 
Since, the WSS model includes an infinite number of vector bosons,
coupled to the pions, it is not necessarily in contradiction with our
results in this paper.
An analysis of what happens to the spectrum as a function of the
number and the masses of the vector bosons could be an interesting,
albeit probably difficult future direction.

\subsection*{Acknowledgments}

S.~B.~G.~thanks the Outstanding Talent Program of Henan University and
the Ministry of Education of Henan Province for partial support.
The work of S.~B.~G.~is supported by the National Natural Science
Foundation of China (Grants No.~11675223 and No.~12071111) and by the
Ministry of Science and Technology of China (Grant No.~G2022026021L).

\bibliographystyle{JHEP}
\bibliography{bib}

\providecommand{\href}[2]{#2}\begingroup\raggedright\begin{thebibliography}{10}

\bibitem{Witten:1998zw}
E.~Witten, {\it {Anti-de Sitter space, thermal phase transition, and
  confinement in gauge theories}},  {\em Adv. Theor. Math. Phys.} {\bf 2}
  (1998) 505--532, [\href{http://arxiv.org/abs/hep-th/9803131}{{\tt
  hep-th/9803131}}].

\bibitem{Sakai:2004cn}
T.~Sakai and S.~Sugimoto, {\it {Low energy hadron physics in holographic QCD}},
   {\em Prog. Theor. Phys.} {\bf 113} (2005) 843--882,
  [\href{http://arxiv.org/abs/hep-th/0412141}{{\tt hep-th/0412141}}].

\bibitem{Rebhan:2014rxa}
A.~Rebhan, {\it {The Witten-Sakai-Sugimoto model: A brief review and some
  recent results}},  {\em EPJ Web Conf.} {\bf 95} (2015) 02005,
  [\href{http://arxiv.org/abs/1410.8858}{{\tt arXiv:1410.8858}}].

\bibitem{Bolognesi:2013nja}
S.~Bolognesi and P.~Sutcliffe, {\it {The Sakai-Sugimoto soliton}},  {\em JHEP}
  {\bf 01} (2014) 078, [\href{http://arxiv.org/abs/1309.1396}{{\tt
  arXiv:1309.1396}}].

\bibitem{Donaldson:1992}
S.~K. Donaldson and P.~B. Kronheimer, {\em {The geometry of four-manifolds}}.
\newblock {Oxford Mathematical Monographs, Clarendon Press}, Oxford, 1990.

\bibitem{Dorey:2002ik}
N.~Dorey, T.~J. Hollowood, V.~V. Khoze, and M.~P. Mattis, {\it {The Calculus of
  many instantons}},  {\em Phys. Rept.} {\bf 371} (2002) 231--459,
  [\href{http://arxiv.org/abs/hep-th/0206063}{{\tt hep-th/0206063}}].

\bibitem{Hitchin:1986ea}
N.~J. Hitchin, A.~Karlhede, U.~Lindstrom, and M.~Rocek, {\it {Hyperkahler
  Metrics and Supersymmetry}},  {\em Commun. Math. Phys.} {\bf 108} (1987) 535.

\bibitem{Atiyah:1968mp}
M.~F. Atiyah and I.~M. Singer, {\it {The Index of elliptic operators. 1}},
  {\em Annals Math.} {\bf 87} (1968) 484--530.

\bibitem{Atiyah:1975jf}
M.~F. Atiyah, V.~K. Patodi, and I.~M. Singer, {\it {Spectral asymmetry and
  Riemannian Geometry I}},  {\em Math. Proc. Cambridge Phil. Soc.} {\bf 77}
  (1975) 43.

\bibitem{Kobayashi:2021jbn}
S.~K. Kobayashi and K.~Yonekura, {\it {The
  Atiyah\textendash{}Patodi\textendash{}Singer index theorem from the axial
  anomaly}},  {\em PTEP} {\bf 2021} (2021), no.~7 073B01,
  [\href{http://arxiv.org/abs/2103.10654}{{\tt arXiv:2103.10654}}].

\bibitem{Matsuki:2021zct}
Y.~Matsuki, H.~Fukaya, M.~Furuta, S.~Matsuo, T.~Onogi, S.~Yamaguchi, and
  M.~Yamashita, {\it {A physicist-friendly reformulation of the mod-two
  Atiyah-Patodi-Singer index}},  {\em PoS} {\bf LATTICE2021} (2022) 617,
  [\href{http://arxiv.org/abs/2111.11040}{{\tt arXiv:2111.11040}}].

\bibitem{Hata:2007mb}
H.~Hata, T.~Sakai, S.~Sugimoto, and S.~Yamato, {\it {Baryons from instantons in
  holographic QCD}},  {\em Prog. Theor. Phys.} {\bf 117} (2007) 1157,
  [\href{http://arxiv.org/abs/hep-th/0701280}{{\tt hep-th/0701280}}].

\bibitem{Atiyah:1978wi}
M.~F. Atiyah, N.~J. Hitchin, and I.~M. Singer, {\it {Selfduality in
  Four-Dimensional Riemannian Geometry}},  {\em Proc. Roy. Soc. Lond. A} {\bf
  362} (1978) 425--461.

\bibitem{Jackiw:1977pu}
R.~Jackiw and C.~Rebbi, {\it {Spinor Analysis of Yang-Mills Theory}},  {\em
  Phys. Rev. D} {\bf 16} (1977) 1052.

\bibitem{Braaten:1985np}
E.~Braaten and L.~Carson, {\it {The Deuteron as a Soliton in the Skyrme
  Model}},  {\em Phys. Rev. Lett.} {\bf 56} (1986) 1897.

\bibitem{Zahed:1986qz}
I.~Zahed and G.~E. Brown, {\it {The Skyrme Model}},  {\em Phys. Rept.} {\bf
  142} (1986) 1--102.

\bibitem{Houghton:1997kg}
C.~J. Houghton, N.~S. Manton, and P.~M. Sutcliffe, {\it {Rational maps,
  monopoles and Skyrmions}},  {\em Nucl. Phys. B} {\bf 510} (1998) 507--537,
  [\href{http://arxiv.org/abs/hep-th/9705151}{{\tt hep-th/9705151}}].

\bibitem{Sutcliffe:2010et}
P.~Sutcliffe, {\it {Skyrmions, instantons and holography}},  {\em JHEP} {\bf
  08} (2010) 019, [\href{http://arxiv.org/abs/1003.0023}{{\tt
  arXiv:1003.0023}}].

\bibitem{Sutcliffe:2011ig}
P.~Sutcliffe, {\it {Skyrmions in a truncated BPS theory}},  {\em JHEP} {\bf 04}
  (2011) 045, [\href{http://arxiv.org/abs/1101.2402}{{\tt arXiv:1101.2402}}].

\bibitem{BjarkeGudnason:2018bju}
S.~B. Gudnason and C.~Halcrow, {\it {Vibrational modes of Skyrmions}},  {\em
  Phys. Rev. D} {\bf 98} (2018), no.~12 125010,
  [\href{http://arxiv.org/abs/1811.00562}{{\tt arXiv:1811.00562}}].

\end{thebibliography}\endgroup
\end{document}